\documentclass[aps,pra,reprint,superscriptaddress,showkeys]{revtex4-2}

\usepackage{graphicx}
\usepackage{amsmath, amssymb}
\usepackage{xcolor}

\begin{document}


\title{Collision and coalescence dynamics of bosonic quantum Hall droplets}


\author{Xinyi Liu}
\affiliation{State Key Laboratory of Information Photonics and Optical Communications, School of Physical Science and Technology, Beijing University of Posts and Telecommunications, Beijing 100876, China}

\author{Zhendong Li}
\affiliation{State Key Laboratory of Information Photonics and Optical Communications, School of Physical Science and Technology, Beijing University of Posts and Telecommunications, Beijing 100876, China}

\author{Yuwen Zhou}
\affiliation{State Key Laboratory of Information Photonics and Optical Communications, School of Physical Science and Technology, Beijing University of Posts and Telecommunications, Beijing 100876, China}

\author{Siying Li}
\affiliation{State Key Laboratory of Information Photonics and Optical Communications, School of Physical Science and Technology, Beijing University of Posts and Telecommunications, Beijing 100876, China}

\author{Haoran Xu}
\affiliation{State Key Laboratory of Information Photonics and Optical Communications, School of Physical Science and Technology, Beijing University of Posts and Telecommunications, Beijing 100876, China}

\author{Zihe Liu}
\affiliation{State Key Laboratory of Information Photonics and Optical Communications, School of Physical Science and Technology, Beijing University of Posts and Telecommunications, Beijing 100876, China}

\author{Rongzhen Jiao}
\affiliation{State Key Laboratory of Information Photonics and Optical Communications, School of Physical Science and Technology, Beijing University of Posts and Telecommunications, Beijing 100876, China}

\author{Mingyuan Sun}
\thanks{Contact author: mingyuansun@bupt.edu.cn}
\affiliation{State Key Laboratory of Information Photonics and Optical Communications, School of Physical Science and Technology, Beijing University of Posts and Telecommunications, Beijing 100876, China}


\date{\today}

\begin{abstract}
Recently bosonic quantum Hall droplets have been observed in rapidly rotating two-dimensional Bose-Einstein condensates (BECs), which exhibit robust dynamical stability. Inspired by this, we systematically investigate the collision and coalescence dynamics of these droplets within the Gross-Pitaevskii framework. For two-droplet collisions, we find two distinct collision outcomes, namely merging and separation, that are controlled by the initial relative velocity. The critical velocity exhibits a universal scaling law with the interaction and the particle number as $v_c \propto (gN)^{1/4}$, which can be interpreted from a simplified analytical model, revealing the essential role of the collision time. It differs fundamentally from the mechanism governing the conventional Lee-Huang-Yang stabilized quantum droplets. Furthermore, while the collision can change the shape of the droplet significantly, the center of mass trajectory remains nearly unaffected, owing to the conservation of angular momentum. For overlapping stationary droplets, vortex arrays can emerge through Kelvin-Helmholtz instability driven by phase-induced shear flow. Although two droplets may merge into a larger one, extended states cannot be constructed from multiple overlapping droplets. Instead, the system dynamically reorganizes into new isolated droplets, revealing the localized property in the bulk region. Our results reveal the unique nonequilibrium dynamics of quantum Hall droplets and suggest new pathways for manipulating strongly correlated rotating quantum fluids. 
\end{abstract}


\maketitle

\section{Introduction}
Bose-Einstein condensates (BECs) have emerged as a versatile platform for exploring a myriad of quantum many-body phenomena since their pioneering realization in dilute atomic gases decades ago \cite{Anderson1995,Davis1995,Bloch2008}. One fascinating frontier within this field is the study of rotating BECs \cite{Wilkin2000,Fetter2009,Li2026}. Under moderate rotation, a quantized vortex lattice can emerge with the breaking of rotational symmetry \cite{AboShaeer2001,Cozzini2006,Williams2010,ORiordan2016,Klaus2022}. In the rapid rotation limit (where the rotational frequency approaches the trapping frequency), the system in the rotating frame can simulate quantum Hall physics \cite{Stormer1999,Cooper2008,Leonard2023}, characterized by incompressible quantum fluids analogous to charged particles in a magnetic field. Recent experiments have found that when a BEC is prepared in a single Landau-gauge wavefunction in the lowest Landau level (LLL), it evolves into a persistent one-dimensional array of quantum Hall droplets instead of a vortex lattice \cite{Mukherjee2022}. Subsequent theoretical work has established the existence of dynamically stable single quantum Hall droplet and droplet array states \cite{Cao2026}. However, nonequilibrium interaction dynamics of these droplets remain largely unexplored. In particular, it is unknown whether colliding droplets merge or separate and whether vortex excitations can emerge during collision or coalescence dynamics. 
 
 The collision and coalescence dynamics have been studied for various systems such as solitons \cite{Zhang2019,Li2023,Zhou2023,Geng2023} and quantum droplets \cite{Zhou2019,Zhou2021,Ferioli2019,AlbaArroyo2022,Hu2022,Adhikari2017,Yang2023,Li2024,Otajonov2024,Hu2025}. In these systems, the scenarios after collision include merging, separation and even evaporation, which are generally governed by collision velocity and particle number. Generally, they are associated with the competition between the kinetic energy and the surface tension. However, various factors can influence the dynamics, depending on the specific system. For example, for mixture droplets, Lee-Huang-Yang (LHY) correction plays an essential role in stabilizing the droplet by offering an effective repulsive interaction and thereby affects the critical velocity \cite{Ferioli2019,AlbaArroyo2022}. For dipolar droplets, the anisotropic nature of dipole-dipole interactions makes the collision outcomes highly sensitive to the collision axis \cite{Adhikari2017}. While these investigations have established a general framework for collisions of self-bound systems, the collisional dynamics of bosonic quantum Hall droplets remains a relatively unknown area. 

In this work, we investigate the collision and coalescence dynamics of quantum Hall droplets in a rapidly rotating BEC by solving the Gross-Pitaevskii equation (GPE). Isolated droplets can maintain stable shapes in this system, with their trajectories governed by their initial velocities \cite{Cao2026}. Therefore, by controlling initial conditions, two identical droplets can be prepared with opposing initial phases such that they follow clockwise circular trajectories and collide with each other (see FIG. \ref{fig:4.1}). Our calculations display a rich dynamical process—interference overlap, merging, and separation, driven by the interplay of velocity and interactions. Then, we explore the coalescence dynamics of overlapping stationary droplet arrays. The superposition of droplets with opposite phases generates a phase gradient, whose associated velocity field leads to the formation of quantized vortices. The vortex number increases as the overlap between droplets increases in some parameter regime. Although two overlapping droplets can coalesce into a larger one, extending this to many droplets does not yield an extended state. Instead, the droplet arrays reorganize into dynamically stable new isolated droplets.

\section{Model}
The rotating quasi-2D BEC can be described by the Gross-Pitaevskii equation (GPE) in the rotating frame as \cite{Gross1961a,Pitaevskii1961,Gross1963}
\begin{equation}
i\hbar \frac{\partial}{\partial t} \psi = \left( -\frac{\hbar^2}{2m} \nabla^2 + V + gN |\psi|^2 - \Omega L_z \right) \psi
\end{equation}
Here, $V = m\omega^2 \left( x^2 + y^2 \right) /2$ is the harmonic trapping potential with the radial frequency $\omega_\perp=\omega$. The term $gN |\psi|^2$ describes the mean‑field interaction with strength $g = a_s \sqrt{8\pi\omega_z /m}$ ($a_s$ is the s-wave scattering length and $\omega_z$ is the $z$-axial trap frequency, making $\omega_z \gg \omega_\perp$ to ensure quasi-2D confinement). $- \Omega L_z$ is the rotational term with angular velocity $\Omega$ and angular momentum operator $L_z = -i \hbar \left( x \partial_y - y \partial_x \right)$. The condensate wave function $\psi$ is normalized to 1. 

In the rapid‑rotation limit where the rotation frequency equals the trap frequency ($\Omega = \omega_\perp$), the centrifugal potential exactly cancels the external harmonic confinement. Adopting the symmetric gauge for the corresponding vector potential $\mathbf{A}(\mathbf{r}) = m \Omega \hat{z} \times \mathbf{r}$, the GPE simplifies to
\begin{equation}
i\hbar \frac{\partial}{\partial t} \psi = \left( -\frac{\hbar^2}{2m} \left( \nabla - i\frac{\mathbf{A}}{\hbar} \right)^2 + gN |\psi|^2 \right) \psi
\end{equation}
This form explicitly highlights the effective magnetic field $\mathbf{B} = \nabla \times \mathbf{A}(\mathbf{r}) = 2m\Omega\hat{z}$. This effective magnetic field gives rise to a cyclotron frequency of $\omega_c = 2\Omega$, so that the period of a complete cyclotron orbit is $T = \pi/\Omega$ (Note that this is half the harmonic oscillator period.). Based on existing experimental conditions \cite{Mukherjee2022}, the radial frequency is set to $\omega_\perp = 2\pi \times 88.6$ Hz, yielding a magnetic length $\ell_B= \sqrt{\hbar/2m\Omega} \approx 1.6$ $\mu$m and a cyclotron period $T \approx 5.6$ ms. The interaction strength parameter is chosen as $gN = 0.0125\omega_c\ell^2_B\times N$, with $N$ being the atom number. 

In the numerical implementation, we adopt dimensionless units. Lengths are scaled by the harmonic oscillator length $\ell_{ho}=\sqrt{\hbar/m\Omega}=\sqrt2\ell_B$ and times are measured in units of $T$. Velocities are in units of $\ell_{ho}\Omega$. For simplification, units are not displayed in the subsequent discussions. Numerical simulations are performed using the time‑splitting Fourier spectral method \cite{Bao2006}. 

Evolving the GPE in imaginary time with zero effective magnetic field yields the steady state of a single quantum Hall droplet \cite{Cao2026}. We take this state as the initial state for our dynamical simulations, and denote it as $\psi_0(\mathbf{r})$. It is centered at the origin and exhibits the characteristic flat‑top density profile and intrinsic clockwise velocity field. Due to the translational invariance established by the magnetic translation theorem, a droplet initially located at an arbitrary position $\mathbf{r}_0$ and carrying a canonical momentum $\mathbf{p}_0$ can be generated via the transformation
\begin{equation}
\psi(\mathbf{r}, 0) =  \psi_0 (\mathbf{r} - \mathbf{r}_0)e^{i \mathbf{p}_0 \cdot \mathbf{r}}
\end{equation}
This construction, which preserves the dynamical stability of the droplet, serves as the starting point for all subsequent studies in this work.

\section{Collision dynamics of two droplets in cyclotron motion}
\subsection{The initial state}
\begin{figure}
\centering
\includegraphics[width=0.48\textwidth]{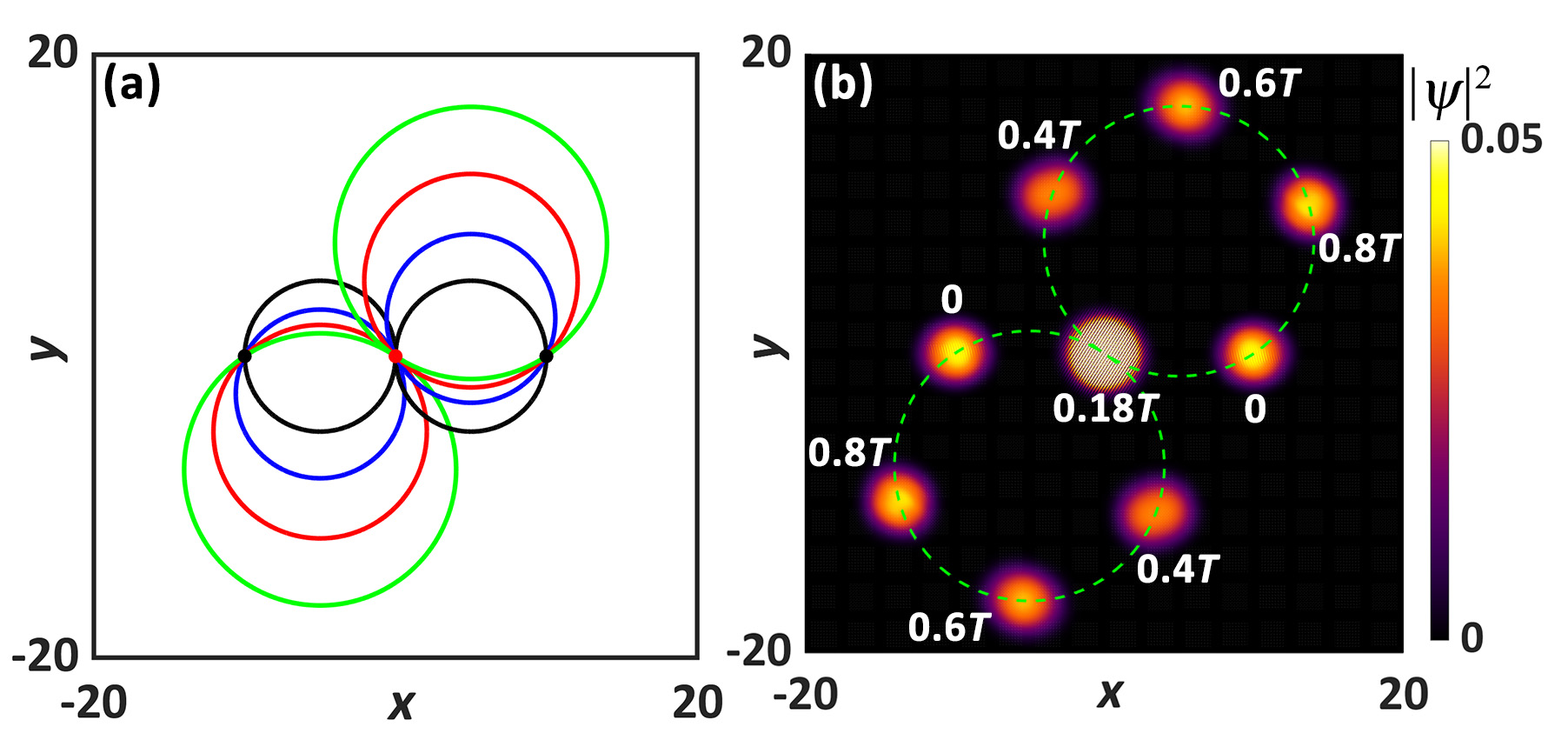}%
\caption{(a) Trajectories of a single droplet initially displaced to the left or right by $L/2 = 10$ for different values of $k$. The trajectories are shown in black, blue, red, and green for $k=0,10,20,30$, respectively. The initial point and the collision point are respectively marked with black dot and red dot. (b) Density snapshots of two colliding droplets at selected times ($0,0.18T,0.4T,0.6T,0.8T$) for the case $k=30$, confirming that the motion of each colliding droplet follows the green circular path as the single-droplet trajectory shown in (a).}
\label{fig:4.1}
\end{figure}

For bosonic quantum Hall droplets in a rapidly rotating BEC, the initial velocity and cyclotron orbit are controllable through phase engineering of the condensate wave function. For instance, a droplet displaced from the trap center without any additional phase acquires an initial velocity perpendicular to the displacement direction due to the magnetic field, resulting in a clockwise circular trajectory passing through the origin. In the following, we study the collision behavior between two identical droplets. The two droplets are prepared by displacing them symmetrically by a distance $L/2$ along the $x$-axis with opposite canonical momenta $\pm k/2$ in the x-direction. Therefore, the initial wave function can be written as
\begin{equation}
\psi = \psi_0 (x + L/2, y) e^{i k x / 2} + \psi_0 (x - L/2, y) e^{-i k x / 2}
\label{initialstate}
\end{equation}
In the rotating frame, the initial mechanical velocity of a droplet at position $\mathbf{r}_0$ with canonical momentum $\mathbf{p}_0$ is
\begin{equation}
\mathbf{v}_0 = \frac{\mathbf{p}_0}{m} - \Omega \hat{\mathbf{z}} \times \mathbf{r}_0
\end{equation}
Hence the left droplet starts with $\mathbf{v}_1(0) = (k/2,\Omega L/2)$ and the right one with $\mathbf{v}_2(0) = (-k/2,-\Omega L/2)$, with the same speed
\begin{equation}
v_0 = \frac{1}{2}\sqrt{k^2 + \Omega^2 L^2}
\end{equation}
The resulting circular orbit has a radius
\begin{equation}
R = \frac{v_0 }{\omega_c} = \frac{1}{4\Omega} \sqrt{k^2 + \Omega^2 L^2}
\end{equation}
and the centers of the orbits are located at $\mathbf{r}_{c1} = (-L/4,- k/(4\Omega))$ and $\mathbf{r}_{c2} = (L/4, k/(4\Omega))$ respectively. A direct calculation shows that the droplets designed with Eq.~\ref{initialstate} always pass through the origin (0,0) simultaneously, where they undergo a head‑on collision, as displayed in FIG.~\ref{fig:4.1}. 

The orientation of the velocities at collision are controlled by the parameters $L$ and $k$. When $k = 0$, the collision occurs at $T/2$ with the velocities perpendicular to the $x$-axis. Increasing k advances the collision and tilts the velocity vectors. The angle between the left droplet’s velocity and the positive $x$-axis at collision is $\theta = \arctan(\Omega L / k)$ which decreases as $k$ grows. In FIG.~\ref{fig:4.1}(b), we demonstrate the evolution of the two droplets with $k=30$ as an example. 

\subsection{Collision dynamics of droplets}
\begin{figure}
\centering
\includegraphics[width=0.48\textwidth]{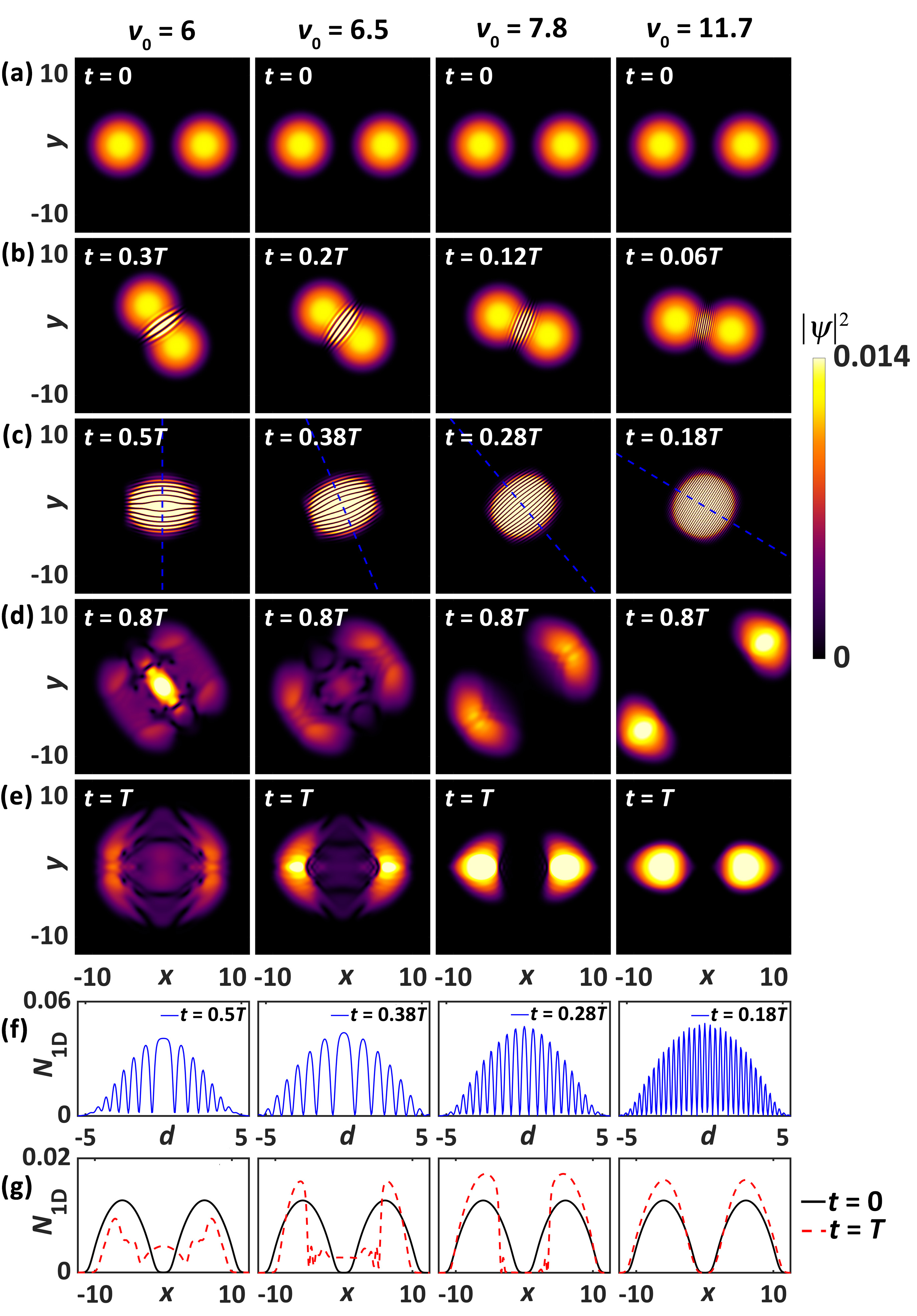}%
\caption{Collision dynamics of droplets for $gN = 1000$. The two identical droplets are initially separated by a distance of $L = 12$ along the $x$-axis and endowed with opposite canonical momenta $\pm k/2$ in the $x$-direction (with dimensionless velocities $v_0=6,6.5,7.8,11.7$ corresponding to $k=0,5,10,20$). Density distribution of the wave function: (a) at the initial state, (b)-(d) during the collision, and (e) after the collision. (f) 1D density $N_{1\mathrm{D}}$ along the axis of symmetry in (c). (g) 1D density $N_{1\mathrm{D}}$ along the line $y = 0$ at $t = 0$ (a) and $t = T$ (e).}
\label{fig:4.2}
\end{figure}
In the following, we investigate the collision dynamics of two identical quantum Hall droplets. As introduced previously, the center‑of‑mass of each droplet performs a uniform cyclotron motion with a period of $T$. Generally speaking, the droplets approach the trap center, gradually overlap, exhibit an interference pattern, pass through each other, and finally return to their starting positions, as displayed in FIG.~\ref{fig:4.2}. A larger $v_0$ leads to a more complete separation after the collision, clearly illustrating the transition from a merged pattern to two separate droplets. Moreover, based on the previously derived collision angle $\theta$, the central symmetry axis of the interference fringe pattern is indicated by the blue dashed line in FIG.~\ref{fig:4.2}(c). The 1D density profile along this axis is plotted in FIG.~\ref{fig:4.2}(f), where the horizontal coordinate d represents the distance along the axis from the center, and the vertical coordinate $N_{1\mathrm{D}}$ denotes the corresponding density. The profile exhibits periodic oscillations whose amplitude is largest at the center and gradually decays toward both ends. This modulation reflects the initial density envelope of a single droplet. With increasing collision speed, the fringes become denser, the peaks become narrower and higher, and the distribution approaches a more regular pattern of equally spaced peaks and valleys. In the non‑interacting (single‑particle) limit, the spacing between adjacent interference peaks is $\lambda=\pi/v_0$. However, the fringe spacings we measure are generally slightly larger than this ideal value, mainly because interactions reduce the effective relative velocity. This deviation is not spatially uniform: closer to the collision center, the density is higher, the effective relative velocity is reduced more, and thus the fringe period is stretched more significantly. Notably, for lower velocities (e.g., $v_0 = 6$), a significantly broadened central peak appears, which prevents the two droplets from fully separating after collision and instead leads to a merged structure in the central region. This low-velocity behavior may be partly attributed to the longer collision time and the reduced collisional kinetic energy, which allow the interaction energy to play a more dominant role. Furthermore, through the 1D density distribution $N_{1\mathrm{D}}(x,t)$ along the line $y = 0$, one can see the droplet profiles undergo a moderate reshaping, characterized by a higher and more concentrated central density (see FIG.~\ref{fig:4.2}(g)). The separation after collision manifests as a near‑zero probability density around $x = 0$ for higher velocities, indicating that the two droplets have moved apart. In addition, a higher collision velocity leads to less deviation of the droplet shape from its initial form, so the separated droplets more closely resemble the original condensates. 
\begin{figure}
\centering
\includegraphics[width=0.45\textwidth]{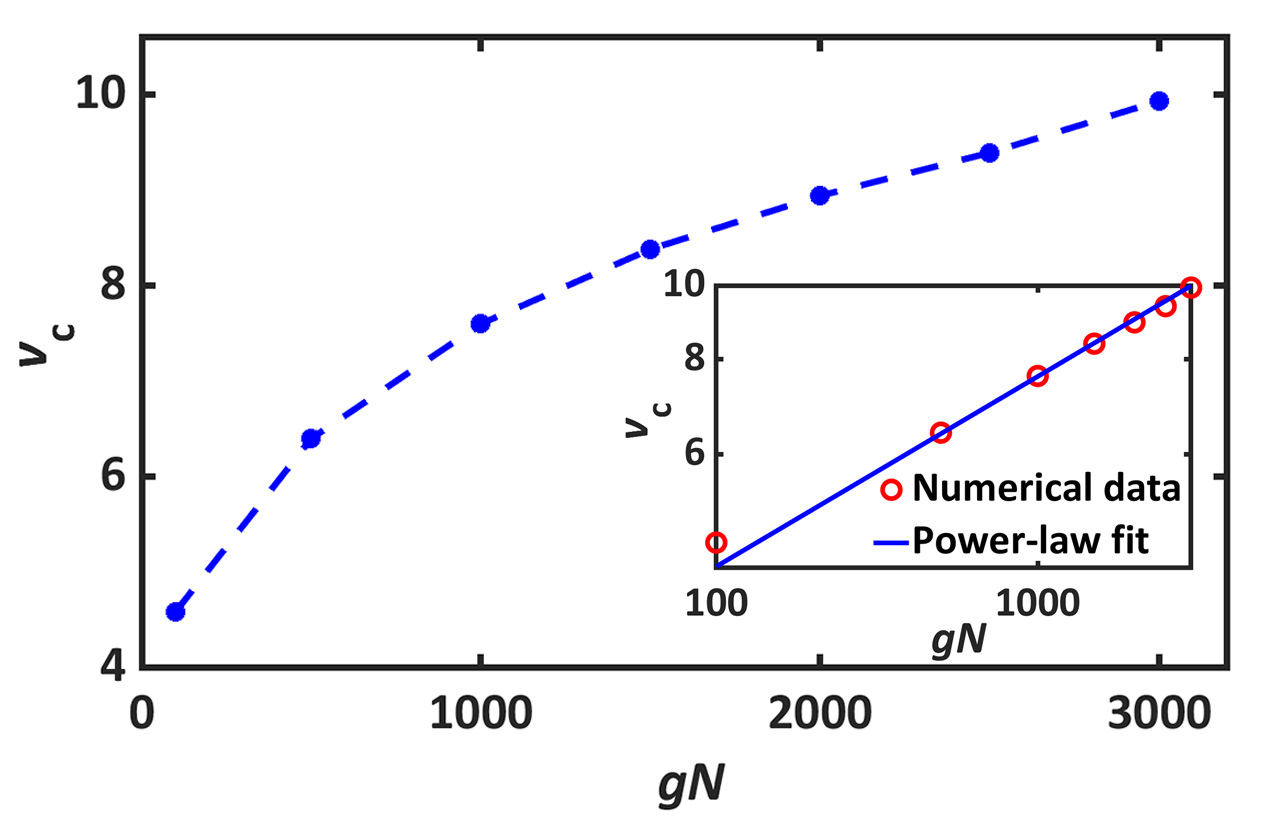}%
\caption{Critical velocity $v_c$ at various $gN$ values. Inset: log‑log power‑law fit, where red circles represent numerical data and the blue solid line is the the fitted line: $\log \hat{v}_c=(1/4)\log (gN)+C$.}
\label{fig:4.3}
\end{figure}
\begin{figure}
\centering
\includegraphics[width=0.48\textwidth]{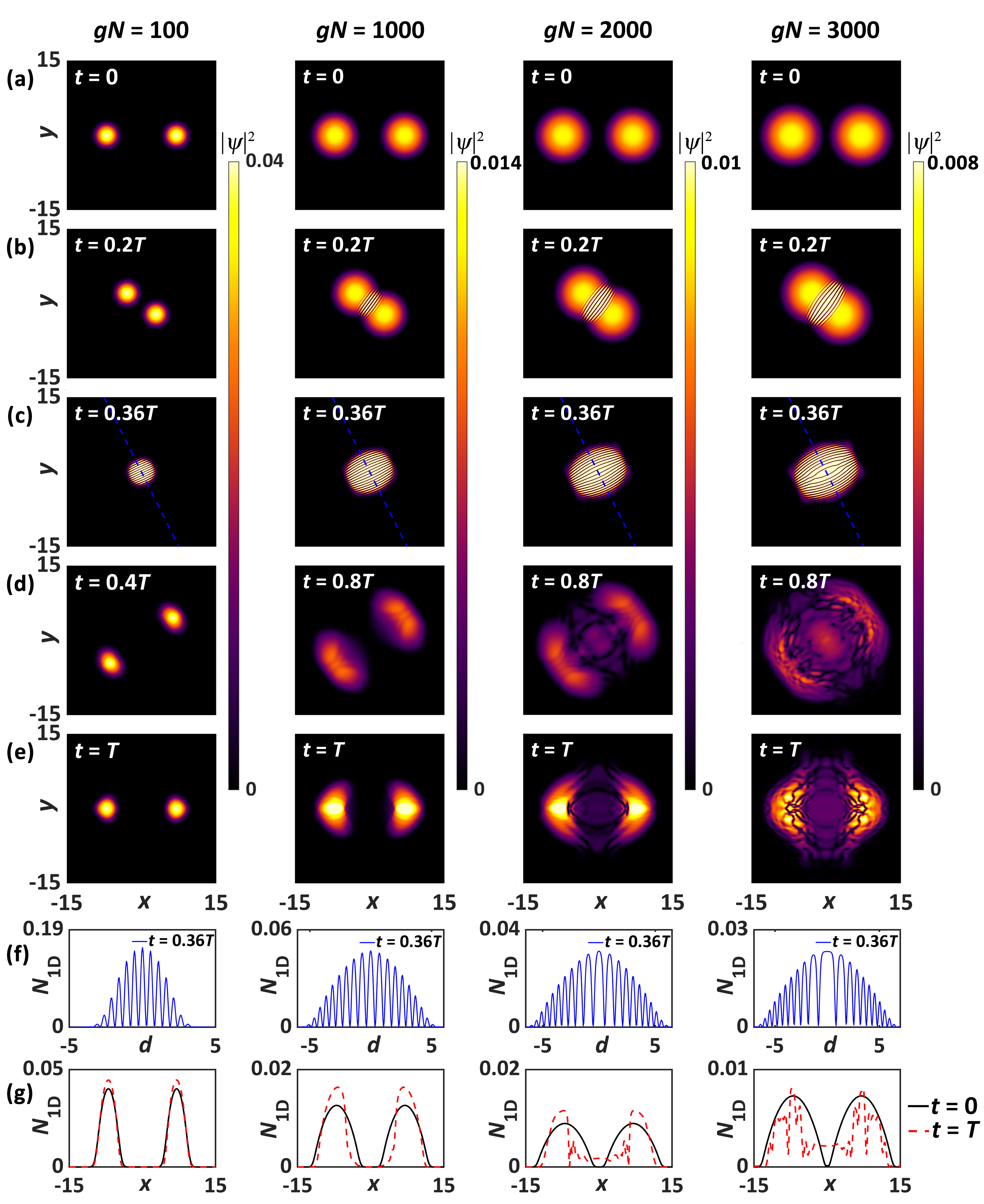}%
\caption{Collision dynamics of droplets for various interaction strengths at fixed initial velocity $v_0 = 7.8$. The interaction strengths are $gN=100,1000,2000,3000$. Density distribution of the wave function: (a) at the initial state, (b)-(d) during the collision, and (e) after the collision. (f) 1D density $N_{1\mathrm{D}}$ along the axis of symmetry in (c). (g) 1D density $N_{1\mathrm{D}}$ along the line y = 0 at $t = 0$ (a) and $t = T$ (e).}
\label{fig:4.4}
\end{figure}

Based on the collision outcomes, the transition between the separation and merging of the droplets is analyzed. To quantify this, we define perfect separation as a state where both droplets return to their initial positions in a uniformly condensed state, with the density along the line $x=0$ falling below $0.05\%$ of the total atom number. The critical velocity $v_c$ is defined as the minimum initial relative velocity of the two droplets at which perfect separation occurs. The obtained $v_c$ as a function of $gN$ is displayed in FIG.~\ref{fig:4.3}. As $gN$ increases, $v_c$ gradually rises. This can be interpreted in a simplified model. Drawing on the Thomas-Fermi approximation, the size of the droplet and the collision time can be written as $R_{TF} \propto (gN)^{1/4}$ and $t_c \propto R_{TF}/v$ respectively. To characterize the collision strength, we introduce the quantity $S=\bar{E_i} t_c$, where $E_i = g |\psi_1|^2 |\psi_2|^2$ is the interaction energy density between the two droplets and $\bar{E_i}$ denotes its average during the collision. Within this estimate, $\bar{E_i}$ can be regarded as a constant approximately. Since $S$ is then proportional to $t_c$, the transition between merging and separation occurs when $S$ reaches a critical value, which gives $R_{TF}/v_c\propto \text{const}$. Therefore, the critical velocity can be expressed as $v_c \propto R_{TF} \propto (gN)^{1/4}$, which agrees well with our numerical results (see inset in FIG.~\ref{fig:4.3}). 

To further explore how the interaction strength influences the collision, we investigate the dynamics with various $gN$ at a fixed initial velocity. FIG.~\ref{fig:4.4} presents the results. It reveals the change in the degree of separation, the interference pattern during overlap, and the final shape of the droplets. Specifically, as $gN$ increases, the outcome changes from nearly perfect separation to a final merged configuration. For weak interactions, the 1D density profile exhibits a nearly periodic oscillation with equally spaced peaks. In contrast, for stronger interactions, the central peak becomes significantly broader and higher, which impedes complete separation and leads to partial or full merging. In line with our previous discussion, this broadening is accompanied by a systematic deviation of the fringe spacing from the ideal value. The deviation grows as the interaction strength increases. This increase is most pronounced near the density maximum and gradually diminishes toward the lower-density edge regions. Similarly, the post‑collision density profile almost coincides with the initial one for weak interactions and the collision is approximately elastic. Motivated by this, we further investigate multiple consecutive collisions. 
\begin{figure}
\centering
\includegraphics[width=0.48\textwidth]{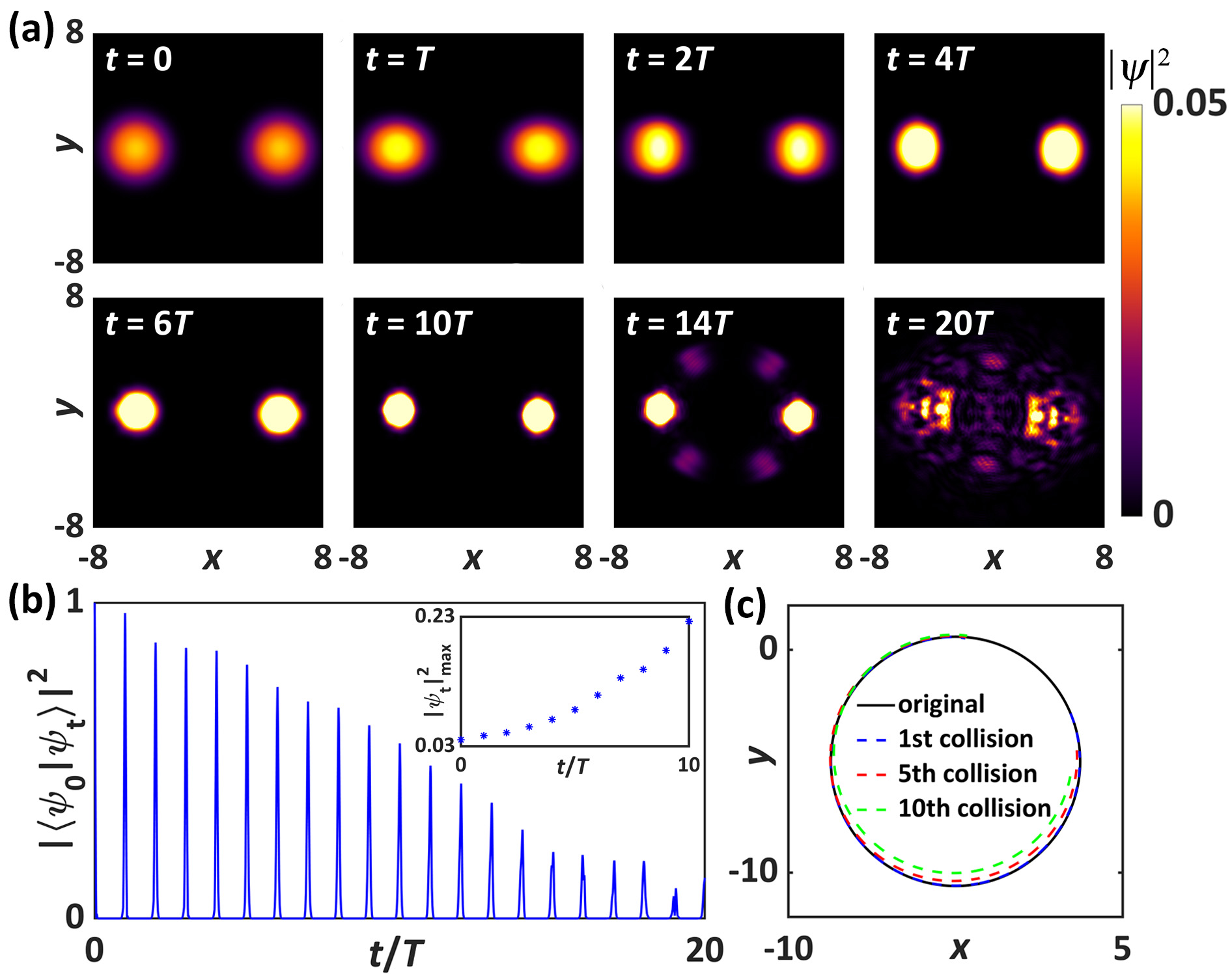}%
\caption{Collision dynamics of droplets for $gN = 100$, $v_0 = 11.2$. (a) Density distribution of the wave function at $t = 0, T, 2T, 4T, 6T, 10T, 14T, 20T$. (b) Survival probability $\left| \langle \psi_0 | \psi_t \rangle \right|^2$ as a function of time $t$. The inset is peak density of the wave function at the end of each cyclotron period. (c) Center-of-mass trajectory of a single droplet. The solid black line represents the dynamics with no collision, while the dashed lines represent the situations after the first, fifth, and tenth collisions.}
\label{fig:4.5}
\end{figure}
\begin{figure}
\centering
\includegraphics[width=0.48\textwidth]{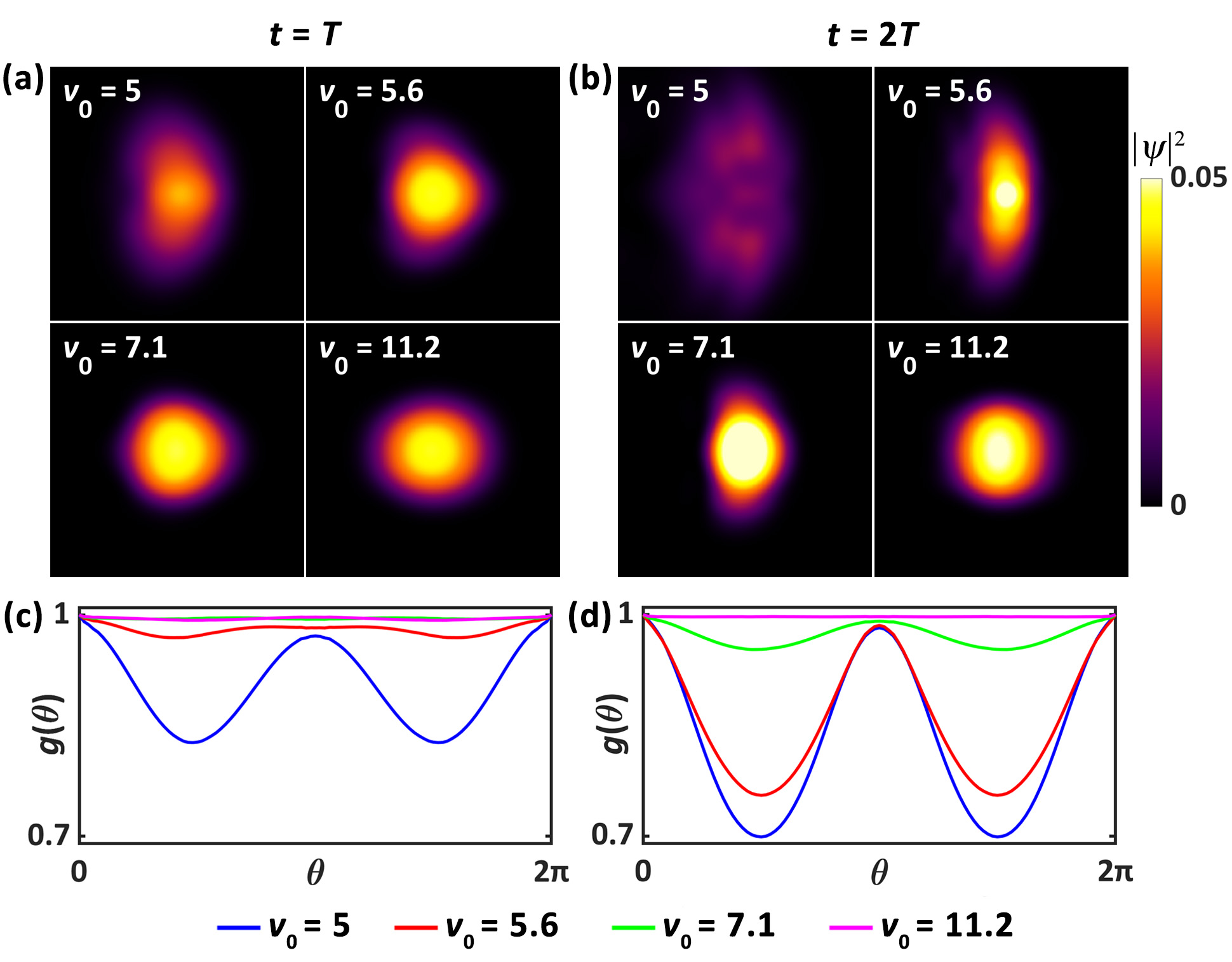}%
\caption{Density distribution and correlation function $g(\theta)$ of the right droplet after collision for different initial velocities $v_0$. Parameters are $gN = 100$ and $L = 10$, with initial velocities $v_0 = 5, 5.6, 7.1, 11.2$. (a) Density and (c) $g(\theta)$ after the first collision ($t=T$). (b) Density and (d) $g(\theta)$ after the second collision ($t=2T$).}
\label{fig:4.6}
\end{figure}

FIG.~\ref{fig:4.5}(a) displays the density distributions after multiple collisions. During the first several cyclotron periods, the droplets remain perfectly separation. Remarkably, as collisions accumulate, each droplet becomes increasingly concentrated. Its core density rises (see inset in FIG.~\ref{fig:4.5}(b)) and spatial extent narrows, resulting in a more compact shape than the initial Gaussian-like profile. Then (take $t = 14T$ as an example), partial fragmentation emerges with the droplets splitting into several smaller density clumps. After many collisions, the structure becomes increasingly complex, exhibiting multiple fragments scattered around the original trajectories. 

To quantify this progressive departure from the initial configuration, we calculate the survival probability, defined as the square modulus of the overlap of the wave functions at time $t$ and the initial one, i.e. $\left| \langle \psi_0 | \psi_t \rangle \right|^2$. The survival probability decreases with each successive collision (see FIG.~\ref{fig:4.5}(b)), implying a steady accumulation of deformations. During the first several collisions, the decay is modest, and the droplets remain highly overlapped with the initial state, consistent with their observed stability and clear separation. As collisions continue, the increasing deviation from the initial wave function marks the progressive reshaping and eventual fragmentation of the droplets. Furthermore, FIG.~\ref{fig:4.5}(c) compares the center-of-mass trajectory of a single droplet after multiple collisions with the cyclotron orbit in the absence of collision. The deviation is negligible compared to the restructuring of the internal density distribution. This contrast underscores the robustness of the cyclotron motion imposed by the effective magnetic field, which persists even as the internal structure of the droplet changes significantly. 

In order to reveal the feature of the internal structure, we calculate the density correlation function for a single droplet \cite{Zhang2020}:
\begin{equation}
g(\theta) = \frac{\int_{0}^{2\pi} d\varphi \,n(\varphi) n(\varphi+\theta)}{\int_{0}^{2\pi} d\varphi \, (n(\varphi))^2}
\end{equation}
Here, $\varphi$ is the angular coordinate and $n(\varphi)$ is the density at that angle. Then $g(\theta)$ measures the correlation as a function of the angular offset $\theta$. We compute the correlation for the right droplet after the first and second collisions (see FIG. \ref{fig:4.6}). After the collision, the droplet exhibits an elongated density profile along the $y$-direction. Its angular correlation function $g(\theta)$ shows a characteristic ``W''-shape, with a peak at $\theta=\pi$ and two symmetric valleys near $\pi/2$ and $3\pi/2$, indicating a two-fold angular modulation. As $v_0$ increases, $g(\theta)$ becomes flatter and the valleys become shallower, suggesting that larger velocity leads to a more isotropic internal structure. 

Our results on binary collisions of bosonic quantum Hall droplets can be compared with the extensive studies on mixture droplets stabilized by LHY correction. In both systems, the collision dynamics exhibits two distinct regimes, namely merging and separation, governed by a critical relative velocity $v_c$. However, the mechanisms differ fundamentally. Mixture droplets rely on the balance between attractive mean‑field interactions and repulsive LHY quantum fluctuations, whereas quantum Hall droplets in the rotating frame are stabilized by the balance between outward mean‑field pressure and the inward Coriolis force. Consequently, the dependence of $v_c$ on droplet size (or atom number) is different. For mixture droplets, $v_c(N)$ is nonmonotonic: it increases with $N$ for small droplets and decreases for large droplets \cite{Ferioli2019,AlbaArroyo2022}. In contrast, for bosonic quantum Hall droplets, $v_c$ increases monotonically with $N$, indicating the incompressible nature of the LLL fluid. Furthermore, the effect of the collision is closely associated with the quantity $\bar{E_i}t_c$, where the collision time also plays an important role. Moreover, additional features such as vortex formation in overlapping stationary arrays are observed, which will be discussed in the following section on coalescence dynamics. These comparisons highlight the distinct role of the effective magnetic field in shaping the collisional behavior of quantum Hall droplets, underscoring their difference from LHY‑stabilized quantum liquids.

\section{Coalescence dynamics of overlapping stationary droplets}
\subsection{The initial state}
\begin{figure*}
\centering  
\includegraphics[width=0.95\textwidth]{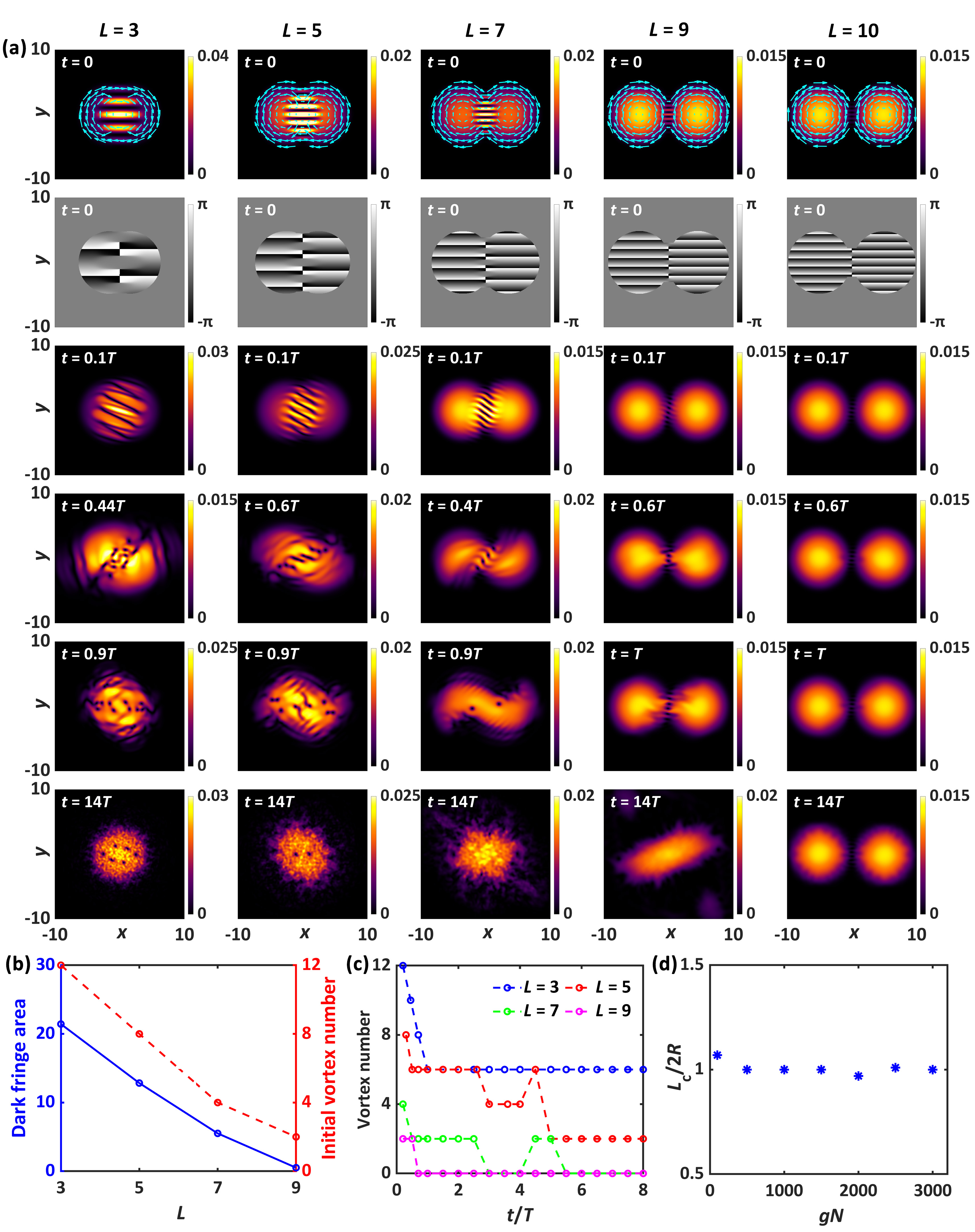}%
\caption{(a) Initial density and phase distributions, with arrows indicating the velocity field, and coalescence dynamics of droplets for $gN = 1000$. The distances between the two droplets are $L=3,5,7,9,10$. Colorbar in color represents density $|\psi|^2$ and grayscale bar represents phase $\theta$. (b) Total area of dark fringes (blue, left axis) and initial vortex number (red, right axis) as functions of $L$. (c) Time evolution of the vortex number. (d) The ratio of the critical separation distance to the droplet diameter $L_c/2R$ as a function of $gN$ ranging form 100 to 3000, demonstrating the threshold between coalescence and separation.}
\label{fig:5.1}
\end{figure*}
The collision below the critical velocity can lead to significant change of the droplet's structure or even new features. Moreover, one related question is, can one connect an array of droplets to form a continuous BEC (an extended state)? Therefore, we investigate the coalescence dynamics of the overlapping stationary droplets. Specifically applying the phase $e^{-iLy}$ to the droplet displaced left by $L$ can make it stationary. In this way, an even-numbered droplet array is formed by symmetrically displacing the central quantum Hall droplet left and right by a distance of $nL/2$ ($n = 1,3,5,…$). Adjusting the inter-droplet spacing $L$ tunes the overlap between droplets. The wavefunction can be expressed as
\begin{equation}
\begin{aligned}
&\psi = \psi_1 \left( x - \frac{L}{2}, y \right) e^{-iLy/2} + \psi_2 \left( x + \frac{L}{2}, y \right) e^{iLy/2} \\
&+ \psi_3 \left( x - \frac{3L}{2}, y \right) e^{-3iLy/2} + \psi_4 \left( x + \frac{3L}{2}, y \right) e^{3iLy/2} + \dots \\
\label{Eqlap}
\end{aligned}
\end{equation}

\subsection{Coalescence dynamics of droplets}
We first study two overlapping droplets. The initial state is prepared by symmetrically displacing the droplets by $L$ along the $x$-axis and applying opposite phases as shown in Eq.~(\ref{Eqlap}), where $L$ controls the initial separation and hence the overlap. The opposite phase factors generate a relative shear flow in the $y$-direction within the overlap region. Studies have shown that such a shear flow leads to a regular array of quantized vortices along the shear layer, due to Kelvin-Helmholtz instability \cite{HernandezRajkov2024}. FIG.~\ref{fig:5.1}(a) depicts the initial density and phase distributions and the subsequent evolution dynamics. The two droplets gradually merge into a unified one and rotate clockwise. The superposition generates interference fringes of the density as
\begin{equation}
|\psi|^2 = |\psi_1|^2 + |\psi_2|^2 + 2|\psi_1||\psi_2|\cos(Ly + \phi)
\end{equation}
with wavelength $\lambda = 2\pi/L$ in the $y$-direction. The velocity field $\mathbf{v}$ reveals a clockwise circulation inside each droplet. In the overlap region, the superposition of these two flow patterns generates small-scale clockwise current structures and induces a rotational motion. 

Over time, the velocity field causes the fringes to gradually tilt, compress, and distort, while the phase develops sharp gradients near the dark regions. As the fringes break and reconnect, isolated zero‑amplitude points emerge, where the local density is dynamically suppressed to zero to avoid divergent kinetic energy. Around such a point, the single‑valuedness of the wave function quantizes the phase circulation as $\oint \nabla \varphi \cdot d\mathbf{l} = 2\pi q\,(q \in \mathbb{Z}_{\neq 0})$ \cite{Klaus2022}, thereby forming quantum vortices with a phase singularity at its core. The generated vortices further rotate with the background flow. FIG. \ref{fig:5.1}(b) displays the dependence of the initial dark-fringe area $A_d$ and the number of initially formed vortices $N_{vi}$ on $L$. From our numerical calculations, the total area of dark fringes is approximately half of the overlap area of the two droplets, each modeled as a circle of radius $R = 4.75$. The overlap area can be calculated using the circle-circle intersection formula $A_o = 2R^2 \arccos(L/2R) -( L/2) \sqrt{4R^2 - L^2}$ for $ L \leq 2R$. As $L$ increases, both the dark‑fringe area and the initial vortex number decrease. Notably, an approximately linear relationship exists between $A_d$ and $N_{vi}$ as shown by our numerical results, leading us to propose that $A_d$ may serve as a geometric indicator for vortex nucleation. 


FIG. \ref{fig:5.1}(c) displays the time evolution of the vortex number $N_v$. Quantized vortex cores are identified by a phase winding of $\pm 2\pi$. To eliminate spurious signals arising from low‑density regions, we consider only vortices with local density exceeding 10\% of the maximum density. Some vortices dissipate upon reaching the outer low‑density regions, leading to a gradual reduction in vortex number. Simulations show that for large overlapping, vortices can persist throughout the evolution. Although $N_v$ continues to decrease slowly, it eventually stabilizes after reaching a certain level. For small overlapping, vortices gradually disappear. When the overlap is too weak for coalescence, the droplets remain in place but develop small disturbances. Therefore, a critical separation distance $L_c$ exists at which the droplets just cease to merge. As shown in FIG. \ref{fig:5.1}(d), the ratio $L_c/2R$ remains approximately 1, indicating that the coalescence of two quantum Hall droplets is primarily determined by geometric overlap rather than interaction strength. It is consistent with our physical interpretation of collision dynamics, where the collision time is infinite for the coalescence case. This geometric criterion is different from the coalescence of classical liquid droplets \cite{Ryu2023,Beaty2023,Eggers2025} and other quantum droplets \cite{Escartin2019,Escartin2022}. 

\begin{figure}
\centering
\includegraphics[width=0.48\textwidth]{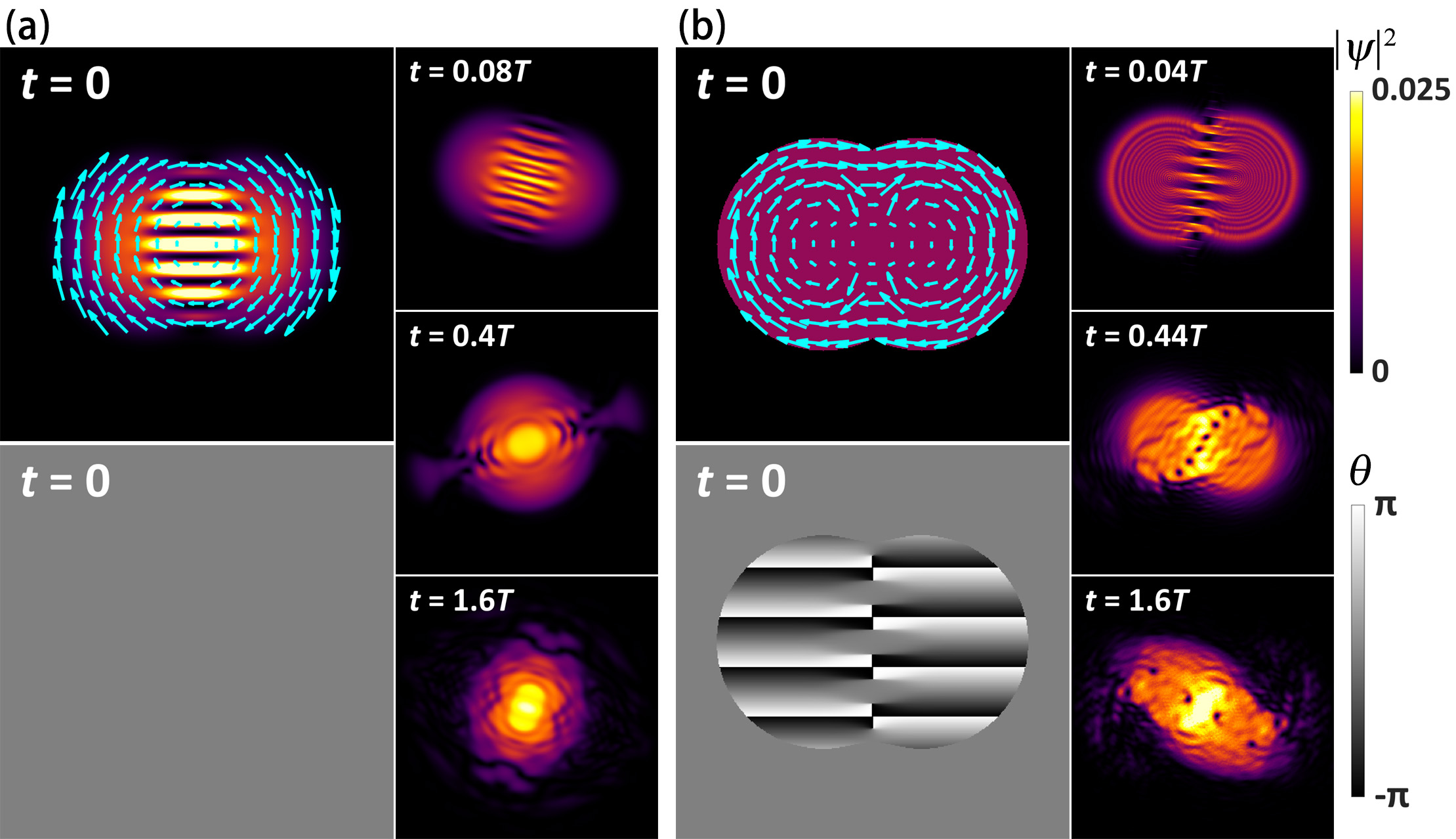}%
\caption{Initial density and phase distributions, with arrows indicating the velocity field, and coalescence dynamics of droplets for $gN = 1000$, $L = 5$ after (a) modulo operation processing and (b) density homogenization.}
\label{fig:5.2}
\end{figure}
\begin{figure}
\centering
\includegraphics[width=0.48\textwidth]{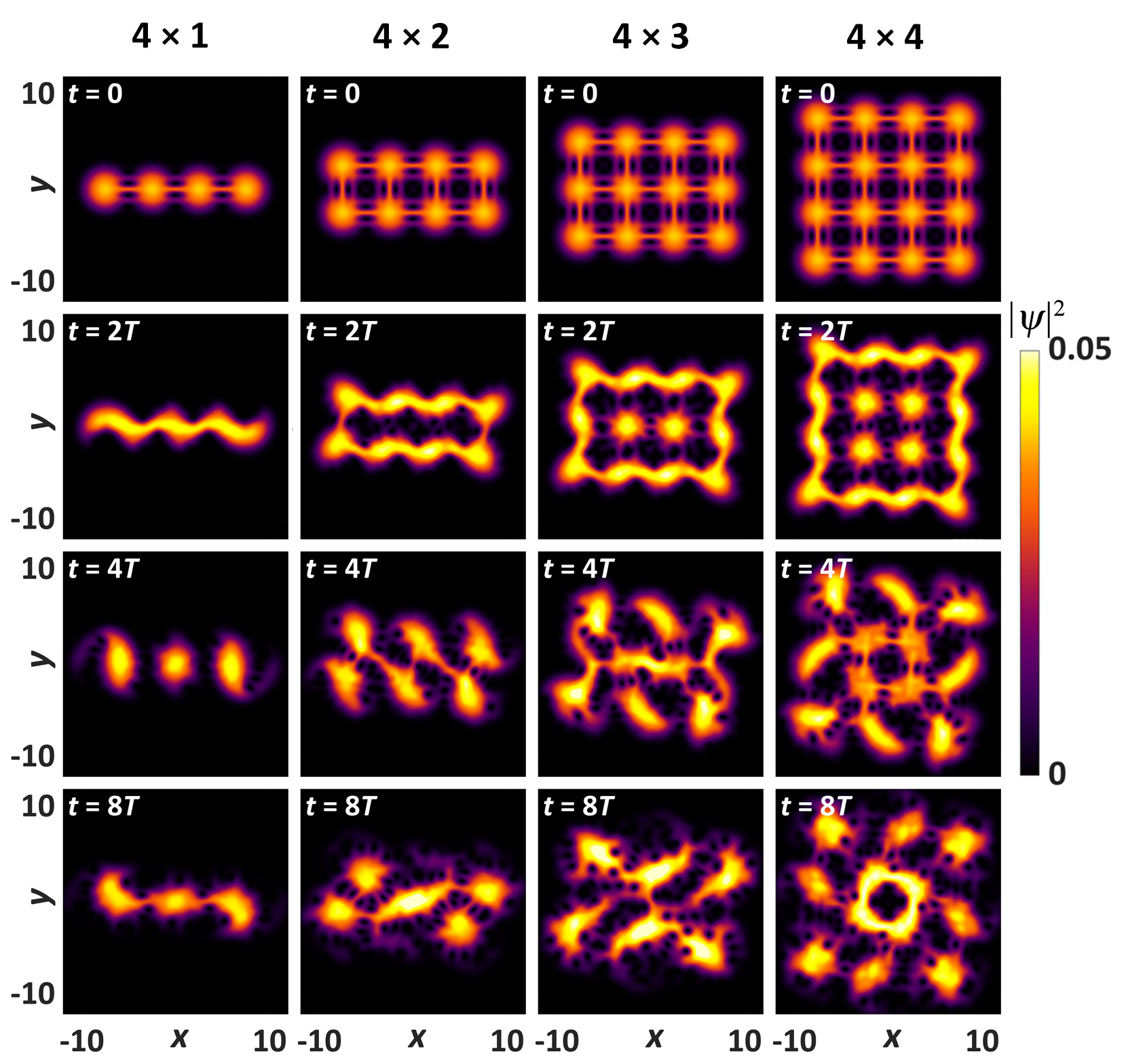}%
\caption{Coalescence dynamics of droplet arrays with each droplet satisfying $gN=50$, for one-dimensional ($4\times 1$) and two-dimensional ($4\times 2,4\times 3,4\times 4$) configurations.}
\label{fig:5.3}
\end{figure}
To further understand the mechanism of vortex formation, we consider two distinct cases (see FIG. \ref{fig:5.2}). The first case is taking the modulus of the superimposed wave function in Eq.~(\ref{Eqlap}) as the initial state, which removes the phase gradient $\nabla \varphi$. The velocity field therefore arises solely from the gauge term. Although the initial interference pattern of the density is preserved, these dark fringes cannot develop quantum vortices. 

The second case is density homogenization. In this setup, we construct a constant density as $\psi = C\frac{\psi_1+\psi_2}{|\psi_1+\psi_2|}$, where $C$ is the normalization constant. This operation preserves the full phase information while removing the density modulation arising from interference. The velocity near each droplet center is nearly zero, while finite within the overlap region, consistent with the initial velocity field in FIG. \ref{fig:5.1}(a). During evolution, density fluctuations emerge due to the Kelvin-Helmholtz instability \cite{Takeuchi2010,Kokubo2021}. The system first develops fringe-like density modulations, which ultimately become unstable and evolve into quantum vortices. These two complementary numerical experiments clearly demonstrate that the velocity field induced by the phase gradient is the fundamental driving force for vortex formation. 

From the above results, one can find that two overlapping droplets generally merge into a larger one at fairly long-time dynamics. Can this conclusion be generalized to systems containing more droplets? In other words, is it possible to construct an extended state through the overlap of many droplets? To answer this question, we investigate the coalescence dynamics of droplet arrays, starting with the one-dimensional case and crossover to two-dimensional configurations. For example, multiple identical droplets are arranged to form larger arrays and their coalescence dynamics are examined as displayed in FIG.~\ref{fig:5.3}. For the one-dimensional case, the initially overlapping droplets gradually merge under rotation and interparticle interactions. They first form a wavy elongated structure and eventually split into a new array of isolated droplets, dynamically stable over long times. Moreover, as the configuration extends to two dimensions, similar dynamical behavior is observed. These results are actually consistent with the general physical picture of quantum Hall effect that the bulk state should be localized in the perpendicular magnetic field. Therefore, our results imply there should be no way to construct an extended state in our system and any connecting droplet array will eventually evolve into separate parts. 

\section{Conclusions}
In conclusion, we systematically investigate the collision and coalescence dynamics of quantum Hall droplets. For two-droplet collisions, varying the collision velocity results in either merging or separation. The critical velocity obeys the scaling law $v_c \propto (gN)^{1/4}$. It can be well interpreted with our model, where the collision time plays an essential role. It is fundamentally different from mixture droplets. For the stationary case, phase-gradient-driven vortices can appear within the overlapping region, due to the Kelvin-Helmholtz instability. Moreover, it may not be possible to construct an extended state with no boundary, agreeing well with the physical picture of quantum Hall effect. These findings not only advance our understanding of quantum Hall droplet dynamics but also provide new insights into emergent properties of strongly correlated quantum systems under rotation, potentially opening new avenues for manipulating topological quantum fluids. 

\begin{acknowledgments}
We thank Zhigang Wu for helpful discussions. This work is supported by NSFC under Grant No. 12374243 (M.Y.S.), BUPT No. 2025JCTP03, Xiaomi Young Talents Program.
\end{acknowledgments}

\bibliographystyle{apsrev4-2}
\bibliography{reference}

@article{Anderson1995,
  author = {Anderson, M. H. and Ensher, J. R. and Matthews, M. R. and others},
  title = {{Observation of Bose-Einstein Condensation in a Dilute Atomic Vapor}},
  journal = {Science},
  volume = {269},
  pages = {198},
  year = {1995}
}

@article{Davis1995,
  author = {Davis, K. B. and Mewes, M. O. and Andrews, M. R. and others},
  title = {{Bose-Einstein Condensation in a Gas of Sodium Atoms}},
  journal = {Phys. Rev. Lett.},
  volume = {75},
  pages = {3969},
  year = {1995}
}

@article{Bloch2008,
  author = {Bloch, Immanuel and Dalibard, Jean and Zwerger, Wilhelm},
  title = {Many-body physics with ultracold gases},
  journal = {Rev. Mod. Phys.},
  volume = {80},
  pages = {885},
  year = {2008}
}

@article{Wilkin2000,
  author = {Wilkin, N. K. and Gunn, J. M. F.},
  title = {Condensation of "composite bosons" in a rotating BEC},
  journal = {Phys. Rev. Lett.},
  volume = {84},
  pages = {6},
  year = {2000}
}

@article{Fetter2009,
  author = {Fetter, A. L.},
  title = {Rotating trapped Bose-Einstein condensates},
  journal = {Rev. Mod. Phys.},
  volume = {81},
  pages = {647},
  year = {2009}
}

@article{Li2026,
  author = {Li, Zhendong and Cao, Huanyu and Cao, Zhen and Li, Siying and Liu, Xinyi and Lan, Yueheng and Sun, Mingyuan},
  title = {Dynamics of the breathing mode with rotational symmetry in two-dimensional Bose-Einstein condensates},
  journal = {Chaos Solitons Fract.},
  volume = {209},
  pages = {118344},
  year = {2026}
}

@article{AboShaeer2001,
  author = {Abo-Shaeer, J. R. and Raman, C. and Vogels, J. M. and others},
  title = {{Observation of Vortex Lattices in Bose-Einstein Condensates}},
  journal = {Science},
  volume = {292},
  pages = {476},
  year = {2001}
}

@article{Williams2010,
  author = {Williams, R. A. and Al-Assam, S. and Foot, C. J.},
  title = {{Observation of Vortex Nucleation in a Rotating Two-Dimensional Lattice of Bose-Einstein Condensates}},
  journal = {Phys. Rev. Lett.},
  volume = {104},
  pages = {050404},
  year = {2010}
}

@article{ORiordan2016,
  author = {O'Riordan, L. J. and Busch, T.},
  title = {{Topological defect dynamics of vortex lattices in Bose-Einstein condensates}},
  journal = {Phys. Rev. A},
  volume = {94},
  pages = {053603},
  year = {2016}
}

@article{Klaus2022,
  author = {Klaus, Lauritz and Bland, Thomas and Poli, Elena and others},
  title = {Observation of vortices and vortex stripes in a dipolar condensate},
  journal = {Nat. Phys.},
  volume = {18},
  pages = {1453-1458},
  year = {2022}
}

@article{Cozzini2006,
  author = {Cozzini, M. and Stringari, S. and Tozzo, C.},
  title = {{Vortex lattices in Bose-Einstein condensates: From the Thomas-Fermi regime to the lowest-Landau-level regime}},
  journal = {Phys. Rev. A},
  volume = {73},
  pages = {023615},
  year = {2006}
}

@article{Stormer1999,
  author = {Stormer, H. L. and Tsui, D. C. and Gossard, A. C.},
  title = {{The fractional quantum Hall effect}},
  journal = {Rev. Mod. Phys.},
  volume = {71},
  pages = {S298},
  year = {1999}
}

@article{Cooper2008,
  author = {Cooper, N. R.},
  title = {{Rapidly rotating atomic gases}},
  journal = {Adv. Phys.},
  volume = {57},
  pages = {539},
  year = {2008}
}

@article{Leonard2023,
  author = {L\'{e}onard, J. and Kim, S. and Kwan, J. and others},
  title = {{Realization of a fractional quantum Hall state with ultracold atoms}},
  journal = {Nature},
  volume = {619},
  pages = {495},
  year = {2023}
}

@article{Mukherjee2022,
  author = {Mukherjee, B. and Shaffer, A. and Patel, P. B. and others},
  title = {{Crystallization of bosonic quantum Hall states in a rotating quantum gas}},
  journal = {Nature},
  volume = {601},
  pages = {58},
  year = {2022}
}

@article{Cao2026,
  author = {Cao, Z. and Li, S. and Li, Z. and others},
  title = {{Bosonic quantum Hall droplets in rapidly rotating two-dimensional Bose-Einstein condensates}},
  journal = {Phys. Rev. A},
  volume = {113},
  pages = {L011301},
  year = {2026}
}

@article{Zhang2019,
  author = {Zhang, Z. and He, Z. and Miao, P. and others},
  title = {{Collision dynamics of two bright solitons in growth Bose-Einstein condensates with tunable interactions}},
  journal = {Int. J. Mod. Phys. B},
  volume = {33},
  pages = {1950261},
  year = {2019}
}

@article{Li2023,
  author = {Li, J. and Yang, Z. and Zhang, S.},
  title = {{Periodic collision theory of multiple cosine-Hermite-Gaussian solitons in Schr\"{o}dinger equation with nonlocal nonlinearity}},
  journal = {Appl. Math. Lett.},
  volume = {140},
  pages = {108588},
  year = {2023}
}

@article{Zhou2023,
  author = {Zhou, Q. and Huang, Z. and Sun, Y. and others},
  title = {{Collision dynamics of three-solitons in an optical communication system with third-order dispersion and nonlinearity}},
  journal = {Nonlinear Dyn.},
  volume = {111},
  pages = {5757},
  year = {2023}
}

@article{Geng2023,
  author = {Geng, K. and Zhu, B. and Cao, Q. and others},
  title = {{Nondegenerate soliton dynamics of nonlocal nonlinear Schr\"{o}dinger equation}},
  journal = {Nonlinear Dyn.},
  volume = {111},
  pages = {16483},
  year = {2023}
}

@article{Ferioli2019,
  author = {Ferioli, G. and Semeghin, G. and Masi, L. and others},
  title = {{Collisions of Self-Bound Quantum Droplets}},
  journal = {Phys. Rev. Lett.},
  volume = {122},
  pages = {090401},
  year = {2019}
}

@article{Hu2022,
  author = {Hu, Y. and Fei, Y. and Chen, X. and others},
  title = {{Collisional dynamics of symmetric two-dimensional quantum droplets}},
  journal = {Front. Phys.},
  volume = {17},
  pages = {61505},
  year = {2022}
}

@article{Hu2025,
  author = {Hu, J. and Wang, H. and Chen, G. and others},
  title = {{The stability and collision dynamics of quantum droplets in PT-symmetric optical lattices}},
  journal = {Chaos Solitons Fract.},
  volume = {191},
  pages = {115837},
  year = {2025}
}

@article{Otajonov2024,
  author = {Otajonov, S. R. and Umarov, B. A. and Abdullaev, F. K.},
  title = {{Dynamics of quasi-one-dimensional quantum droplets in Bose-Bose mixtures}},
  journal = {Chaos Solitons Fract.},
  volume = {186},
  pages = {115212},
  year = {2024}
}

@article{Zhou2019,
  author = {Zhou, Z. and Yu, X. and Zou, Y. and others},
  title = {{Dynamics of quantum droplets in a one-dimensional optical lattice}},
  journal = {Commun. Nonlinear Sci. Numer. Simul.},
  volume = {78},
  pages = {104881},
  year = {2019}
}

@article{Zhou2021,
  author = {Zhou, Z. and Shi, Y. and Tang, S. and others},
  title = {{Controllable dissipative quantum droplets in one-dimensional optical lattices}},
  journal = {Chaos Solitons Fract.},
  volume = {150},
  pages = {111193},
  year = {2021}
}

@article{AlbaArroyo2022,
  author = {Alba-Arroyo, J. E. and Caballero-Benitez, S. F. and J{\'a}uregui, R.},
  title = {Weber number and the outcome of binary collisions between quantum droplets},
  journal = {Sci. Rep.},
  volume = {12},
  pages = {18467},
  year = {2022},
  doi = {10.1038/s41598-022-22904-8}
}

@article{Adhikari2017,
  author = {Adhikari, S. K.},
  title = {Statics and dynamics of a self-bound dipolar matter-wave droplet},
  journal = {Laser Phys. Lett.},
  volume = {14},
  pages = {025501},
  year = {2017},
  doi = {10.1088/1612-202X/14/2/025501}
}

@article{Yang2023,
  author = {Yang, A. and Li, G. and Jiang, X. and Fan, Z. and Chen, Z. and Liu, B. and Li, Y.},
  title = {Two-Dimensional Quantum Droplets in Binary Dipolar Bose-Bose Mixture},
  journal = {Photonics},
  volume = {10},
  number = {4},
  pages = {405},
  year = {2023}
}

@article{Li2024,
  author = {Li, G. and Jiang, X. and Liu, B. and others},
  title = {{Two-dimensional anisotropic vortex quantum droplets in dipolar Bose-Einstein condensates}},
  journal = {Front. Phys.},
  volume = {19},
  pages = {22202},
  year = {2024}
}

@article{Gross1961a,
  author = {Gross, E. P.},
  title = {{Structure of a quantized vortex in boson systems}},
  journal = {Nuovo Cim.},
  volume = {20},
  pages = {454},
  year = {1961}
}

@article{Pitaevskii1961,
  author = {Pitaevskii, L. P.},
  title = {{Vortex lines in an imperfect bose gas}},
  journal = {J. Exp. Theor. Phys.},
  volume = {13},
  pages = {451},
  year = {1961}
}

@article{Gross1963,
  author = {Gross, E. P.},
  title = {{Hydrodynamics of a superfluid condensate}},
  journal = {J. Math. Phys.},
  volume = {4},
  pages = {195},
  year = {1963}
}

@article{Bao2006,
  author = {Bao, W. and Wang, H.},
  title = {{An efficient and spectrally accurate numerical method for computing dynamics of rotating Bose-Einstein condensates}},
  journal = {J. Comput. Phys.},
  volume = {217},
  pages = {612},
  year = {2006}
}

@article{Zhang2020,
  author = {Zhang, Z. and Yao, K. and Feng, L. and others},
  title = {Pattern formation in a driven Bose-Einstein Condensate},
  journal = {Nat. Phys.},
  volume = {16},
  pages = {652--656},
  year = {2020}
}

@article{HernandezRajkov2024,
  author = {Hernandez-Rajkov, D. and Grani, N. and Scazza, F. and others},
  title = {{Connecting shear-flow and vortex array instabilities in annular atomic superfluids}},
  journal = {Nat. Phys.},
  volume = {20},
  pages = {939},
  year = {2024}
}

@article{Ryu2023,
  author = {Ryu, Sangjin and Zhang, Haipeng and Anuta, Udochukwu John},
  title = {A Review on the Coalescence of Confined Drops with a Focus on Scaling Laws for the Growth of the Liquid Bridge},
  journal = {Micromachines},
  volume = {14},
  number = {11},
  pages = {2046},
  year = {2023}
}

@article{Beaty2023,
  author = {Beaty, Edward and Lister, John R.},
  title = {Inertial and viscous dynamics of jump-to-contact between fluid drops under van der Waals attraction},
  journal = {J. Fluid Mech.},
  volume = {957},
  pages = {A25},
  year = {2023}
}

@article{Eggers2025,
  author = {Eggers, Jens and Sprittles, James E. and Snoeijer, Jacco H.},
  title = {Coalescence Dynamics},
  journal = {Annu. Rev. Fluid Mech.},
  volume = {57},
  pages = {61--87},
  year = {2025}
}

@article{Escartin2019,
  title = {Vorticity and quantum turbulence in the merging of superfluid Helium nanodroplets},
  author = {Escart\'in, Jos\'e Mar\'ia and Ancilotto, Francesco and Barranco, Manuel and Pi, Mart\'i},
  journal = {Phys. Rev. B},
  volume = {99},
  pages = {140505(R)},
  year = {2019}
}

@article{Escartin2022,
  title = {Merging of superfluid helium nanodroplets with vortices},
  author = {Escart\'in, J. M. and Ancilotto, F. and Barranco, M. and Pi, M.},
  journal = {Phys. Rev. B},
  volume = {105},
  pages = {024511},
  year = {2022}
}

@article{Takeuchi2010,
  author = {Takeuchi, H. and Suzuki, N. and Kasamatsu, K. and Saito, H. and Tsubota, M.},
  title = {Quantum Kelvin-Helmholtz instability in phase-separated two-component Bose-Einstein condensates},
  journal = {Phys. Rev. B},
  volume = {81},
  pages = {094517},
  year = {2010}
}

@article{Kokubo2021,
  author = {Kokubo, H. and Kasamatsu, K. and Takeuchi, H.},
  title = {Pattern formation of quantum Kelvin-Helmholtz instability in binary superfluids},
  journal = {Phys. Rev. A},
  volume = {104},
  pages = {023312},
  year = {2021}
}

\end{document}